\documentclass[fleqn,usenatbib]{mnras}

\usepackage{newtxtext,newtxmath}
\usepackage[T1]{fontenc}

\DeclareRobustCommand{\VAN}[3]{#2}
\let\VANthebibliography\thebibliography
\def\thebibliography{\DeclareRobustCommand{\VAN}[3]{##3}\VANthebibliography}

\usepackage{graphicx}	% Including figure files
\usepackage{amsmath}	% Advanced maths commands
\usepackage{url}

\title[Scotogenic WIMP expectations]{MHz to TeV expectations from scotogenic WIMP dark matter}

\author[Eisenberger, Siegert, Mannheim, Porod]{
	Laura Eisenberger,$^{1}$\thanks{E-mail: laura.eisenberger@stud-mail.uni-wuerzburg.de}
	Thomas Siegert,$^{1}$\thanks{E-mail: thomas.siegert@uni-wuerzburg.de}
	Karl Mannheim$^{1}$
	and Werner Porod$^{2}$
	\\
	$^{1}$Julius-Maximilians-Universität Würzburg, Fakultät für Physik und Astronomie, Institut für Theoretische Physik und Astrophysik,\\ Lehrstuhl für Astronomie, Emil-Fischer-Str. 31, D-97074 Würzburg, Germany\\
	$^{2}$Julius-Maximilians-Universität Würzburg, Fakultät für Physik und Astronomie, Institut für Theoretische Physik und Astrophysik,\\ Emil-Hilb-Weg 22, D-97074 Würzburg, Germany
}

\date{Accepted XXX. Received YYY; in original form ZZZ}

\pubyear{2023}

\begin{document}
	\label{firstpage}
	\pagerange{\pageref{firstpage}--\pageref{lastpage}}
	\maketitle
	
	\begin{abstract}
		% maximum 250 words!
		%
		The indirect search for dark matter is typically restricted to individual photon bands and instruments.
		In the context of multiwavelength observations, finding a weak signal in large fore- and backgrounds at only one wavelength band is hampered by systematic uncertainties dominating the signal strength.
		Dark matter particle annihilation is producing Standard Model particles of which the prompt photon emission is searched for in many studies.
		However, also the secondary emission of charged particles from dark matter annihilation in the TeV range results in comparable or even stronger fluxes in the GHz--GeV range.
		
		\noindent In this study, we calculate the prompt and secondary emission of a scotogenic WIMP with a mass of $1$\,TeV in 27 dwarf galaxies of the Milky Way.
		For the secondary emission, we include Inverse Compton scattering, bremsstrahlung, and synchrotron radiation, which results in a `triple hump' structure characteristic for only dark matter and no other astrophysical source.
		In order to determine the best candidates for multi-instrument analyses, we estimate the diffuse emission component of the Milky Way itself, including its own dark matter halo from the same scotogenic WIMP model.
		We find signal-to-background ratios of individual sources on the order of $10^{-3}$--$10^{-2}$ across X- to $\gamma$-rays assuming $J$-factors for the cold dark matter distribution inferred from observations and no additional boosting due to small-scale clumping.
		We argue that a joint multi-wavelength analysis of all nearby galaxies as well as the extension towards the Cosmic Gamma-ray Background is required to disentangle possible dark matter signals from astrophysical back- and foregrounds.
	\end{abstract}
	
	
	%%%%%%%%%%%%%%%%%%%%%%%%%%%%%%%%%%%%%%%%%%%%%%%%%%
	
	%%%%%%%%%%%%%%%%% BODY OF PAPER %%%%%%%%%%%%%%%%%%
	
	\section{Introduction}\label{sec:intro}
	The dark matter (DM) phenomenon is supposedly responsible for different astrophysical, cosmological, and particle physics measurements.
	However, decades of searches -- direct or indirect -- for the possible particle character of DM have only yielded null results.
	This is at odds with the `WIMP miracle' \citep{Jungman1996_susydm} for which supersymmetric particles beyond the Standard Model of particle physics (SM) are predicted at the weak scale.
	
	While the `standard' WIMPs may not be the candidate to be looking for, the scotogenic WIMP has attracted more attention again recently \citep{Alvarez_2023}.
	Scotogenic models implement an additional symmetry under which all SM particles are even while new particles are odd.
	This symmetry automatically leads to a stable DM candidate and the generation of neutrino masses through couplings to the dark sector \citep[e.g.,][]{Ma2006_scotogenic,Avila2020_scotoWIMP}.
	Furthermore, some scotogenic models can explain, at the same time, the muon anomalous magnetic moment and, potentially, the baryon asymmetry of the Universe while fulfilling the current limits for charged lepton flavour violating processes \citep{Alvarez_2023}.
	One such candidate is a fermionic WIMP with a mass around 1\,TeV.
	
	Astrophysical, that is, indirect searches for TeV-scale DM has been performed in several studies including Fermi/LAT \citep[e.g.,][]{Ackermann2014_FermiLAT_dsphs,Ackermann2017_GeVGCE}, MAGIC \citep[e.g.][]{MAGIC2016_DM}, or HESS \citep[e.g.,][]{Abdalla2022_HESS_DM} for electromagnetic signals, and PAMELA \citep[e.g.,][]{Adriani2013_PAMELA} or AMS-02 \citep[e.g.,][]{Aguilar2013_AMS} for cosmic-ray excesses at Earth.
	Most of these studies are prone to search for the prompt emission of DM annihilation (or decay) signals, and only recently a few studies included the possible secondary emission from Inverse Compton scattering of leptons off the Cosmic Microwave Background radiation or bremsstrahlung \citep{Cirelli2009_IC_darkmatter,Cirelli2021_subGeVDM,Cirelli2023_subGeVDM,Saxena2011_ICdarkmatter,Djuvsland2023_ICDM}, for example.
	Multiwavelength observations of the prompt emission together with the seconadary and tertiary emission suggest much better chances to observe DM signatures coherently instead of only the prompt signal.
	Secondary emission stems from the stable SM particles originating in the (scotogenic) WIMP annihilation:
	electrons and positrons will undergo Inverse Compton scattering, bremsstrahlung, and synchrotron radiation.
	Tertiary emission might then be considered the annihilation of secondary particles leading to yet another very distinct emission of propagated and cooled positrons, leading to both annihilation in flight and Positronium formation \citep{Siegert2016_511,Siegert2022_RetII}.
	Because the electron propagation from TeV to eV energies is hardly understood in diverse interstellar environments and conditions, such as in the Milky Way \citep{Siegert2023_511}, we focus this work on the prompt and secondary emission from dwarf galaxies which are mostly devoid of any remaining gas.

	This paper is structured as follows:
	In Sect.\,\ref{sec:scotoWIMP}, we describe the scotogenic WIMP model in detail.
	Sect.\,\ref{sec:multiwavelength} includes the expected photon emission from GHz to GeV, considering primary and secondary effects.
	We compute a list of expected fluxes owing to DM for different instruments for all our 27 considered galaxies in Sect.\,\ref{sec:dwarf_expectations}.
	The astrophysical fore- and background emission in these bands are discussed in Sect.\,\ref{sec:astro_foreground}.
	Finally, we discuss our findings in terms of detectability with future instrumentation in the X- to $\gamma$-ray regime in Sect.\,\ref{sec:discussion}.

	%This is a simple template for authors to write new MNRAS papers.
	%See \texttt{mnras\_sample.tex} for a more complex example, and \texttt{mnras\_guide.tex}
	%for a full user guide.
	
	%All papers should start with an Introduction section, which sets the work
	%in context, cites relevant earlier studies in the field by \citet{Fournier1901},
	%and describes the problem the authors aim to solve \citep[e.g.][]{vanDijk1902}.
	%Multiple citations can be joined in a simple way like \citet{deLaguarde1903, delaGuarde1904}.

	\section{Scotogenic WIMP}\label{sec:scotoWIMP}
	Scotogenic models implement an additional $\mathbb{Z}_{2}$ symmetry under which all Standard Model particles are even while new particles are odd.
	This symmetry automatically leads to a stable DM candidate and the generation of neutrino masses through couplings to the dark sector.
	In our study, we use a scotogenic model which can additionally explain the muon anomalous magnetic moment and, potentially, the baryon asymmetry of the Universe while fulfilling the current limits for charged lepton flavour violating (cLFV) processes as \citet{Alvarez_2023} has shown.
	In this framework, the SM is extended by two fermion doublets, two fermion singlets, a scalar doublet, and a scalar singlet.
	The model parameters are the respective new couplings in the scalar potential, the Yukawa couplings, the masses and the lightest neutrino mass.
	The couplings of the new fields to the SM leptons are constrained by neutrino oscillation data and cLFV upper limits and are fitted in a way that they reproduce the experimental value of $(g-2)_{\mu}$.

	The model has three types of possible DM candidates: the lightest dark neutral fermion, the lighter scalar or the pseudo-scalar.
	Further constrained by the Higgs mass $m_\text{H}=(125.3\pm1.0)$\,GeV and DM relic density $\Omega_{\text{CDM}}h^{2}=0.120\pm0.012$, a model parameter scan of \citet{Alvarez_2023} showed that fermionic DM with a mass around 1.1\,TeV dominates the parameter space.
	The spin-independent direct cross section of all fermionic DM candidates is far below the experimental limits meaning that there are no further constraints.
	For this reason, we choose for the following analysis a viable point of the parameter space resulting in a fermionic WIMP DM type with $m_{\chi}=1048.759$\,GeV.

	\section{Primary and secondary emission from annihilating scotogenic WIMP dark matter}\label{sec:multiwavelength}
	The expected $\gamma$-rays are produced both by primary emission from annihilating DM and by secondary processes from charged particles generated during annihilation including Inverse Compton scattering, bremsstrahlung, and synchrotron radiation.
	The differential flux of a particle species $a$ produced directly by annihilation of Majorana DM particles and measured from the direction $\psi$ is
	\begin{equation}
		\frac{d\Phi_{a}}{dE}=\frac{1}{4\pi} \frac{\langle \sigma v \rangle}{2m_{\chi}^{2}} \frac{dN_{a}}{dE} \times \frac{1}{\Delta \Omega} \int_{\Delta \Omega} d\Omega \int_{\text{los}} \rho^{2}(\psi, l)\, dl\,\mathrm{,}
		\label{eq:flux}
	\end{equation}
	where $\langle \sigma v \rangle$, $m_{\chi}$ and $dN_{a}/dE$ are the thermally-averaged annihilation cross section, the mass of the DM particle, and the distribution of the produced particle $a$ per annihilation, respectively.
	The second part of Eq.\,(\ref{eq:flux}) is referred to as the $J$-factor and incorporates the astrophysical dependence.
	It consists of the line-of-sight integral over the squared DM density $\rho$, averaged over the chosen solid angle $\Delta \Omega$.

	\subsection{Primary/prompt emission}\label{sec:primary_emission}
	DM particles annihilate into pairs of SM particles which hadronize and finally decay into stable SM particles (prompt emission).
	Interesting final states for indirect DM searches are $\gamma$-rays, electrons/positrons and neutrinos.
	While the annihilation spectra of photons and neutrinos can be directly converted into observable fluxes according to Eq.\,(\ref{eq:flux}) (primary emission), the propagation of charged particles is governed by diffusion and energy losses and dependent on magnetic fields.
	However, the relativistic electrons and positrons will lead to secondary photons as soon as they are produced (see Sec.\,\ref{sec:secondary_emission}).
	The annihilation cross section and the direct annihilation spectra $dN_{a}/dE$ depend on the respective particle model.
	For their calculation, we use \texttt{micrOMEGAs}, a code written in $C$ to compute WIMP properties for direct and indirect detection given certain particle interactions \citep{belanger2014micromegas}.
	The annihilation cross section was determined by \texttt{micrOMEGAs} to be $\langle \sigma v \rangle = 9.52 \times 10^{-27}\,\text{cm}^{3}\,\text{s}^{-1}$ for the chosen DM candidate with $m_{\chi}=1.05$\,TeV.

\begin{figure}
		\centering
		\includegraphics[width=\columnwidth]{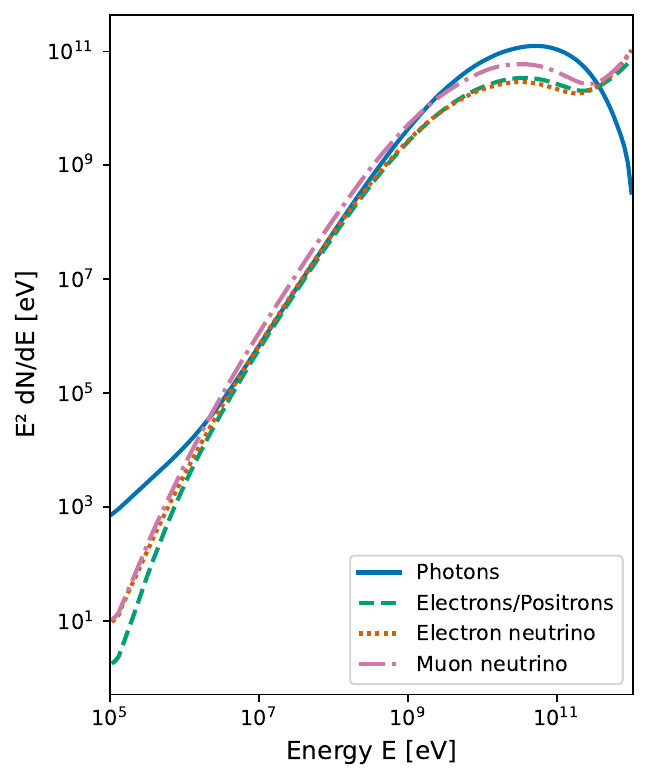}
		\caption{Annihilation spectra computed by \texttt{micrOMEGAs} showing the distributions of different particles produced per annihilation in the scotogenic model described in Sec.\,\ref{sec:scotoWIMP}.} 
		\label{fig:dNdE}
	\end{figure}

	Fig.\,\ref{fig:dNdE} shows the distributions of the above mentioned particles per annihilation for the scotogenic model described in Sec.\,\ref{sec:scotoWIMP}.
	We note that the electron and positron spectra are identical.
	The computation of the annihilation spectra is limited to a minimum energy of $E_{\text{min}}=10^{-7}E_{\text{max}}$, where $E_{\text{max}}$ is the mass of the annihilating DM particle.
	All spectra exhibit a (local) maximum at around 50\,GeV, which corresponds to 5\% of the DM mass and is expected for pion decays after hadronization.
	We note that no line feature was calculated for the photon spectrum since direct annihilation into $\gamma$-rays is loop-suppressed.
	Electrons and electron-neutrinos are produced at the same amount by charged pion decays leading to similar spectra, while twice as many muon neutrinos are created per pion decay.
	
	\subsection{Secondary emission}\label{sec:secondary_emission}
	Charged particles produced by DM annihilation will interact with their environment and thereby create $\gamma$-rays via different mechanisms.
	Highly energetic electrons and positrons will Inverse Compton scatter off the low-energy photons of the interstellar radiation field consisting of approximately three blackbody spectra of different temperatures:
	the CMB, the infrared (IR) dust emission and the optical starlight (SL), where $T_{\text{CMB}}=2.753$\,K, $T_{\text{IR}}=3.5\,$meV and $T_{\text{SL}}=0.3$\,eV \citep[e.g.,][]{Cirelli_2009}.
	While the CMB photon field is present everywhere to the same amount, the IR and SL depends on the respective environment.
	Due to the low content of dust in dwarf spheroidal galaxies, the IR photon field can be neglected when determining the DM IC component.
	The upscattered SL photon field depends on the visual luminosity and size of the respective source.
	Bremsstrahlung is produced when relativistic particles interact with another particle population and therefore depends on the particle density of the respective surroundings.
	Since dwarf galaxies are probably devoid of interstellar medium, the DM contribution to bremsstrahlung in the further vicinity of the Milky Way halo is expected to be sub-dominant but here calculated for completeness.
	Under presence of magnetic fields, charged particles are deflected and thus emit synchrotron radiation.
	In dwarf galaxies, where the magnetic field might be only by factors larger than that of the intergalactic medium, the synchrotron emission is expected to be sub-dominant but nevertheless relevant for a multiwavelength analysis.

	In order to determine the secondary emission, the total number of electrons/positrons $n_{\text{e}}$ per unit energy needs to be calculated.
	Neglecting diffusion, the loss equation is solved by
	\begin{equation}
		n_{\text{e}}(E) = \frac{1}{b(E)} \int_{E}^{m_{\chi}} Q_{\text{e}}(\tilde{E})\, d\tilde{E}\,\mathrm{,}
		\label{eq:ne}
	\end{equation}
	where $E$ and $b(E)$, respectively, are the electron energy and the total rate of electron energy loss considering inverse Compton, synchrotron and bremsstrahlung losses.
	For a point source at distance $d$, the luminosity source term of DM annihilation is given by
	\begin{equation}
		Q_{\text{e}}(E) = \frac{d\Phi_{\text{e}}(E)}{dE} \times 4\pi d^{2}\,\mathrm{,}
		\label{eq:diff_luminosity}
	\end{equation}
	where the differential flux $d\Phi_{\text{e}}/dE$ for electrons/positrons is given by Eq.\,(\ref{eq:flux}).
	The particle distribution $n_{\text{e}}$ is then used as input to calculate the spectra of inverse Compton scattering, synchrotron radiation and bremsstrahlung, respectively.
	We use the \texttt{Naima} package \citep{zabalza2015naima} to perform these calculations.
	The energy losses of all three processes are summed up in the function $b(E)$ in Eq.\,(\ref{eq:ne}).
	As the produced electrons are mostly relativistic, the following formulas are presented in their high-energy limits.

	The synchrotron and inverse Compton losses \citep{Aharonian,Khangulyan} are both given by
	\begin{equation}
		b(E)_{\,\text{IC/Syn}} = \frac{4}{3} \sigma_{\text{T}} c U_{\text{rad/B}} \Bigl{(} \frac{E}{E_{0}} \Bigr{)}^{2}\,\mathrm{,}
		\label{eq:energy_losses_ICSYN}
	\end{equation}
	where $\sigma_{\text{T}}$, $U_{\text{rad/B}}$ and $E_{0}$ are the Thomson cross section, the energy density of the respective field, and the electron rest energy, respectively.
	The energy losses of electron-electron bremsstrahlung \citep{Baring} are calculated by
	\begin{equation}
		b(E)_{\,\text{e-e}} = \frac{e^{6}n_{0} \ln{(192)}}{12 \pi^{3}\epsilon_{0}^{3} \hbar E_{0}} \frac{E}{E_{0}}\,\mathrm{,}
		\label{eq:BR_losses}
	\end{equation}
	where $n_{0}$ is the density of the electron gas.

	\subsubsection{Inverse Compton scattering}\label{sec:ICS}
	The CMB with a temperature of $T=2.7$\,K and a radiation energy density of $U_{\text{CMB}}=0.26\,\text{eV} \text{cm}^{-3}$ is the most relevant photon field for inverse Compton scattering of electrons in dwarf galaxies.
	The SL radiation field can be estimated from the absolute visual magnitude $M_{\text{V}}$ of a source and its half-light radius $r$.
	The energy density is approximately given by
	\begin{equation}
		U_{\text{SL}} = \frac{f_{\rm V}}{4 \pi c r^{2}} L_{\sun} \times 10^{0.4 (M_{\rm V, \sun}-M_{\text{V}})}\,\mathrm{,}
		\label{eq:rad_density}
	\end{equation}
	where a scaling factor $f_{\rm V}$ takes into account the fact that the approximate blackbody spectrum extends outside of the visual range.
	Given the single data points for the visual magnitude, and therefore luminosity, in dwarf galaxies, it is difficult to estimate $f_{\rm V}$ directly.
	However, since dwarf spheroidal galaxies are characterized by an older stellar population, the temperature of the SL photon field is estimated to be around 4000\,K, which limits $f$ to within an order of magnitude.
	Given the V-band filter is almost Gaussian-shaped with a peak around 540\,nm and a FWHM of 40\,nm, we can estimate that the flux of a 4000\,K blackbody included in the V-band is about 5--10\%.
	Because the average temperature is also uncertain, we allow $f_{\rm V}$ to vary between $0.01$ and $1.0$ in our estimates for the SL inverse Compton contribution.
	The \texttt{Naima} package uses the functions $F_{3}(x_{0})$ and $F_{4}(x_{0})$ (see Eqs.\,(15) and (16) of \citet{Khangulyan}) as analytical approximation for the interaction rate of charged particles with an isotropic blackbody target photon field of temperature $T$:
	\begin{equation}
		\Bigl{(} \frac{dN}{d\omega\, dt} \Bigr{)}_{\text{\,IC}} = \frac{2e^{4}E_{0} \kappa T^{2}}{\pi \hbar^{3}c^{2}E^{2}}
		\times \Bigl{[} \frac{z^{2}}{2(1-z)} F_{3}(x_{0}) + F_{4}(x_{0}) \Bigr{]}\,\mathrm{,}
		\label{eq:IC_spec}
	\end{equation}
	where $x_{0}=\frac{z}{4ET(1-z)}$ and $z=\frac{\omega}{E}$ is the ratio of photon and electron energy.
	The dilution factor $\kappa$ is smaller than $1$ in the case of greybody radiation and is calculated by
	\begin{equation}
		\kappa = \frac{c U_{\text{rad}}}{4 \sigma_{\text{SB}} T^{4}}\,\mathrm{.}
		\label{eq:kappa}
	\end{equation}
	Finally, the differential spectrum is given by
	\begin{equation}
		\frac{d\Phi}{d\omega} = \frac{1}{4\pi d^{2}} \int_{E_{\text{min}}}^{E_{\text{max}}}
		2 n_{\text{e}}(E) \times \frac{dN}{d\omega\, dt}\, dE\,\mathrm{,}
		\label{eq:diffSpec}
	\end{equation}
	where $E_{\text{min}}$ is the lower bound of the electron spectrum shown in Fig.\,\ref{fig:dNdE} and $E_{\text{max}}$ is the mass $m_{\chi}$ of the DM particle.
	The factor of $2$ takes into account the equal amounts of positrons and electrons produced by DM annihilation.
	
	\begin{figure}
		\centering
		\includegraphics[width=\columnwidth,trim=0.0cm 0.0cm 0.0cm 0.0cm, clip=True]{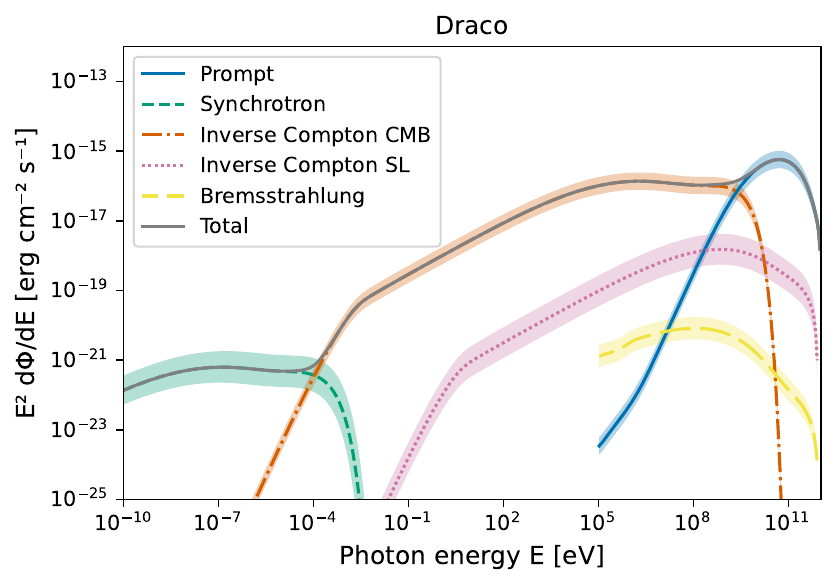}
		\caption{Multiwavelength dark matter spectrum. Shown are the expectations from the direction of the Draco dwarf spheroidal galaxy of prompt emission of dark matter co-annihilation, and the secondary emission of electrons and positrons from the dark matter annihilation undergoing inverse Compton scattering off the CMB and stellar light (SL), bremsstrahlung, and synchrotron radiation. The coloured bands indicate the uncertainties from taking into account the minimum and maximum values of magnetic fields, densities, etc., from modelling the individual components.} 
		\label{fig:spec_Draco}
	\end{figure}

	\begin{table*}
		\begin{tabular}{|c|c|c|c|c|c|c|}
			\hline
			Galaxy & log$_{10}$J (GeV\,cm$^{-5}$) & Distance (kpc) & Magnitude & Radius (pc) & RA (deg) & DEC (deg) \\
			\hline
			Carina & $18.03^{+0.34}_{-0.34}$ & $105\pm6$ & $-9.1\pm0.5$ & $250.0^{+39.0}_{-39.0}$ & 100.40 & -50.97 \\
			
			Draco & $18.92^{+0.25}_{-0.25}$ & $76\pm6$ & $-8.8\pm0.3$ & $221.0^{+19.0}_{-19.0}$ & 260.05 & 57.92 \\
			
			Fornax & $18.27^{+0.17}_{-0.17}$ & $147\pm12$ & $-13.4\pm0.3$ & $710.0^{+77.0}_{-77.0}$ & 40.00 & -34.45 \\
			
			Leo I & $17.80^{+0.28}_{-0.28}$ & $254\pm15$ & $-12.0\pm0.3$ & $251.0^{+27.0}_{-27.0}$ & 152.12 & 12.31 \\
			
			Leo II & $17.41^{+0.25}_{-0.25}$ & $233\pm14$ & $-9.8\pm0.3$ & $176.0^{+42.0}_{-42.0}$ & 168.37 & 22.15 \\
			
			Sculptor & $18.73^{+0.29}_{-0.29}$ & $86\pm6$ & $-11.1\pm0.5$ & $28.3^{+45.0}_{-45.0}$ & 15.04 & -33.71 \\
			
			Sextans & $18.04^{+0.29}_{-0.29}$ & $86\pm4$ & $-9.3\pm0.5$ & $695.0^{+44.0}_{-44.0}$ & 153.26 & -1.61 \\
			
			Ursa Minor & $19.18^{+0.24}_{-0.24}$ & $76\pm3$ & $-8.8\pm0.5$ & $181.0^{+27.0}_{-27.0}$ & 227.29 & 67.22 \\
			
			Boötes I & $16.64^{+0.64}_{-0.38}$ & $66\pm2$ & $-6.3\pm0.2$ & $242.0^{+21.0}_{-21.0}$ & 210.03 & 14.50 \\
			
			Coma Berenices & $18.64^{+0.32}_{-0.32}$ & $44\pm4$ & $-4.1\pm0.5$ & $77.0^{+10.0}_{-10.0}$ & 186.75 & 23.90 \\
			
			Canes Venatici I & $17.27^{+0.11}_{-0.11}$ & $218\pm10$ & $-8.6\pm0.2$ & $564.0^{+36.0}_{-36.0}$ & 202.01 & 33.56 \\
			
			Canes Venatici II & $17.63^{+0.40}_{-0.40}$ & $160\pm4$ & $-4.9\pm0.5$ & $74.0^{+14.0}_{-14.0}$ & 187.83 & 44.05 \\
			
			Hercules & $16.79^{+0.45}_{-0.45}$ & $133\pm12$ & $-6.6\pm0.4$ & $330.0^{+75.0}_{-52.0}$ & 252.78 & 4.99 \\
			
			Leo IV & $16.56^{+0.90}_{-0.90}$ & $154\pm6$ & $-5.8\pm0.4$ & $206.0^{+37.0}_{-37.0}$ & 173.24 & -0.53 \\
			
			Leo V & $16.82^{+1.05}_{-0.70}$ & $178\pm10$ & $-5.2\pm0.4$ & $135.0^{+32.0}_{-32.0}$ & 172.79 & 2.22 \\
			
			Leo T & $17.28^{+0.37}_{-0.37}$ & $417\pm19$ & $-8.0\pm0.5$ & $120.0^{+9.0}_{-9.0}$ & 143.72 & 17.05 \\
			
			Segue 1 & $19.39^{+0.39}_{-0.39}$ & $23\pm2$ & $-1.5\pm0.8$ & $29.0^{+8.0}_{-5.0}$ & 151.77 & 16.08 \\
			
			Segue 2 & $17.06^{+0.86}_{-1.75}$ & $35\pm2$ & $-2.5\pm0.3$ & $35.0^{+3.0}_{-3.0}$ & 34.82 & 20.18 \\
			
			Ursa Major I & $18.47^{+0.25}_{-0.25}$ & $97\pm4$ & $-5.5\pm0.3$ & $319.0^{+50.0}_{-50.0}$ & 158.72 & 51.92 \\
			
			Ursa Major II & $19.38^{+0.39}_{-0.39}$ & $32\pm4$ & $-4.2\pm0.6$ & $149.0^{+21.0}_{-21.0}$ & 132.88 & 63.13 \\
			
			Willman 1 & $19.29^{+0.91}_{-0.62}$ & $38\pm7$ & $-2.7\pm0.8$ & $25.0^{+6.0}_{-6.0}$ & 162.34 & 51.05 \\
			
			Reticulum II & $18.72^{+0.85}_{-0.32}$ & $30\pm3$ & $-2.7\pm0.1$ & $32.0^{+1.9}_{-1.1}$ & 53.93 & -54.05 \\
			
			Tucana II & $19.10^{+0.88}_{-0.58}$ & $57\pm5$ & $-3.8\pm0.1$ & $165.0^{+27.8}_{-18.5}$ & 342.98 & -58.57 \\
			
			Horologium I & $18.64^{+0.95}_{-0.39}$ & $79\pm7$ & $-3.4\pm0.1$ & $30.0^{+4.4}_{-3.3}$ & 43.88 & -54.12 \\
			
			Hydra II & $16.56^{+0.87}_{-1.85}$ & $134\pm10$ & $-4.8\pm0.3$ & $68.0^{+11.0}_{-11.0}$ & 185.43 & -31.99 \\
			
			Pisces II & $17.90^{+1.14}_{-0.80}$ & $182\pm18$ & $-5.0\pm0.0$ & $58.0^{+99.0}_{-99.0}$ & 344.63 & 5.95 \\
			
			Grus I & $17.96^{+0.90}_{-1.93}$ & $120\pm11$ & $-3.4\pm0.3$ & $62.0^{+29.8}_{-13.6}$ & 344.18 & -50.16 \\
			
			\hline
		\end{tabular}
		\caption{List of selected dwarf spheroidal galaxies in this study. $J$-factors are given for an integration angle between the centre and the outermost star of the dwarf. For $J$-factors and distances see \citet{Evans_2016}. Absolute visual magnitudes and half-light radii are taken from \citet{McConnachie_2012}, \citet{Koposov_2015} and \citet{Martin_2015}.} 
		\label{tab:dwarfs}
	\end{table*}

	\subsubsection{Synchrotron radiation}\label{sec:SYN}
	The magnetic field strengths of dwarf spheroidals, that is, such sources which are suitable for DM search, are hardly known.
	However, one can assume that the magnetic field of the Milky Way prevails at the location of its satellites and that it decreases with distance cubed.
	This provides a lower limit on the magnetic field strength at a location in the Galactic halo. 
	The magnetic field strength in our local neighbourhood is about $6\,\mu$G \citep{Beck_2011}.
	Thus, the magnetic field $B$ at distance $d$ and its energy density can be estimated by
	\begin{equation}
		B(d) = 6\, \mu \text{G}\, \Bigl{(} \frac{8\, k\text{pc}}{d} \Bigr{)}^{3}\,\,\,\,\, \text{and}\,\,\,\,\, 
		U_{\text{B}} = \frac{B^{2}}{2\mu_{0}}\,\mathrm{.}
		\label{eq:biot_savart}
	\end{equation}
	For distant satellites, an intergalactic magnetic field strength of 1\,$n$G \citep{Dolag1999} is used as an absolute lower boundary for $B$.

	We again use the \texttt{Naima} package to calculate the spectrum.
	It uses the function $\tilde{G}(x)$ (see Eq.\,(D7) of \citet{Aharonian}) to approximate the synchrotron emissivity averaged over the directions of magnetic field $B$:
	\begin{equation}
		\Bigl{(} \frac{dN}{d\omega\, dt} \Bigr{)}_{\text{\,Syn}} = \frac{\sqrt{3}}{2\pi} \frac{e^{3}B}{E_{0}\hbar \omega} \tilde{G}\Bigl{(} \frac{\omega}{E_{\text{c}}} \Bigr{)}\,,\,\, \text{where}\,\,\, 
		E_{\text{c}} = \frac{3e\hbar cB}{2E_{0}} \Bigl{(} \frac{E}{E_{0}} \Bigr{)}^{2}\mathrm{.}
		\label{eq:synchrotron}
	\end{equation}
	The differential spectrum is then again given by Eq.\,(\ref{eq:diffSpec}).

	\subsubsection{Bremsstrahlung}\label{sec:BR}
	As the gas densities of dwarf galaxies are also hardly known, the particle density of the MW halo scaled with distance is used to estimate the amount of gas in the MW satellites.
	The best-fit spherical model according to \citet{Miller_2013} yields, for large radii, an electron  density of
	\begin{equation}
		n_{0}(d) = 0.46\,\text{cm}^{-3} \Bigl{(} \frac{d}{0.35\,k\text{pc}} \Bigr{)}^{-2.13}\,\mathrm{.}
		\label{eq:density_model}
	\end{equation}
	\texttt{Naima} uses the approximations of \citet{Baring} for the cross-section $\sigma_{\text{e-e}}(E)$ of electron-electron bremsstrahlung which is given by Eqs.\,(A1) and (A5) for the relativistic ($E>2$\,MeV) and non-relativistic case.
	The differential bremsstrahlung spectrum is then given by
	\begin{equation}
		\Bigl{(} \frac{d\Phi}{d\omega} \Bigr{)}_{\text{\,e-e}} = \frac{cn_{0}}{4\pi d^{2}} \int_{E_{\text{min}}}^{E_{\text{max}}}
		2 n_{\text{e}}(E) \times \sigma_{\text{e-e}}(E)\, dE\,\mathrm{.}
		\label{eq:bremsstrahlung}
	\end{equation}

	\section{Expectations in dwarf galaxies}\label{sec:dwarf_expectations}

	Dwarf galaxies typically have large mass-to-light ratios \citep{Strigari2008_dm_dsph}.
	The fact that the faint satellites of the MW must be DM dominated and the low astrophysical backgrounds make them promising targets for the indirect DM search.
	Dwarf spheroidal galaxies (dSphs) are very faint objects having a spheroidal shape and possibly a cusped DM density profile.
	\citet{Evans_2016} analytically determined the $J$-factors for $27$ dSphs by assuming spherical NFW cusps.
	In Tab.\,\ref{tab:dwarfs}, we summarise important properties of the $27$ selected dwarf galaxies.
	The $J$-factors are given for an integration angle between the centre and the outermost star of the dwarf.

\begin{figure*}
		\centering
		\includegraphics[width=0.9\columnwidth]{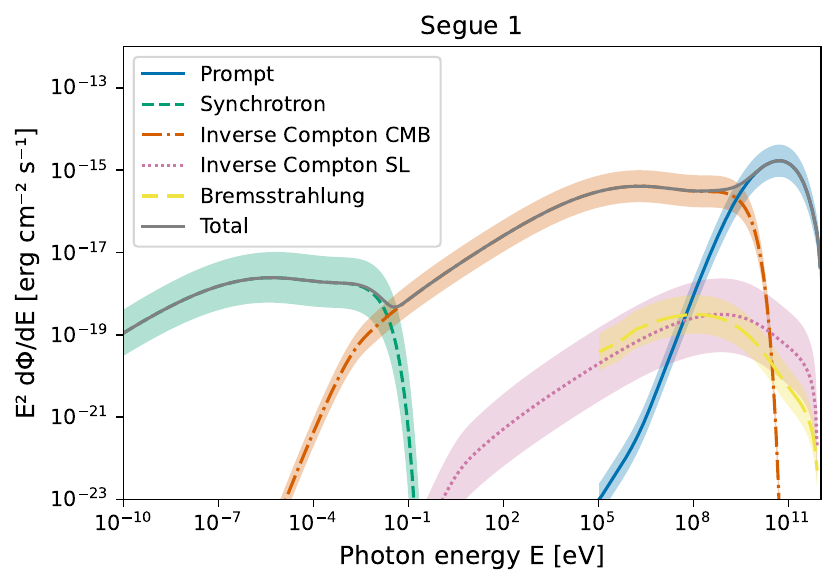}~
		\includegraphics[width=0.9\columnwidth]{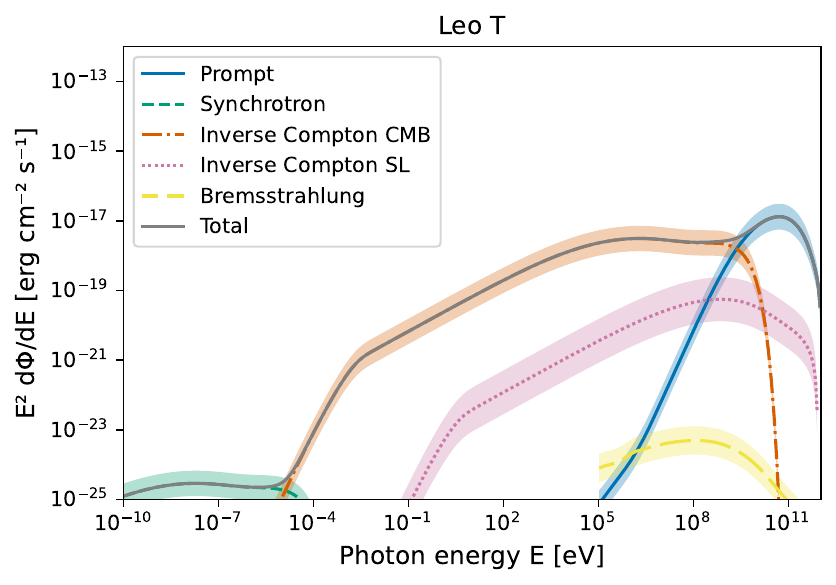}\\
		\includegraphics[width=0.9\columnwidth]{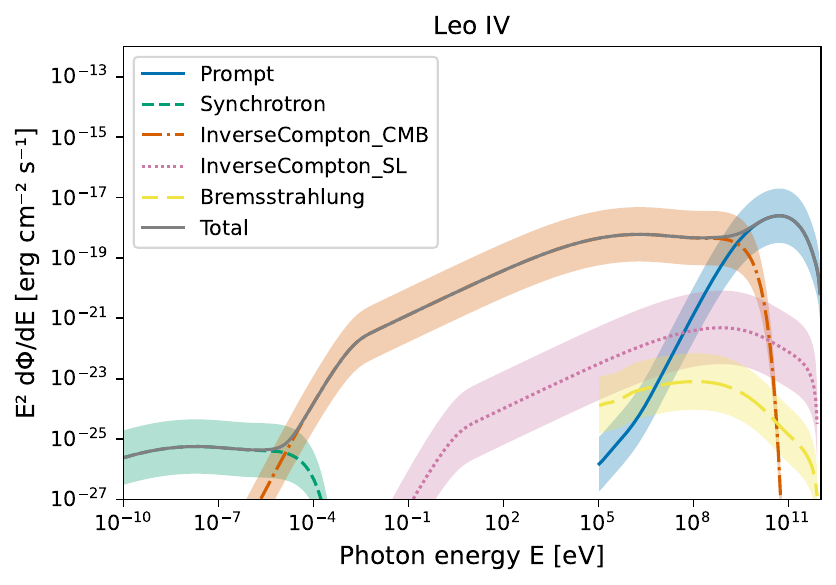}~
		\includegraphics[width=0.9\columnwidth]{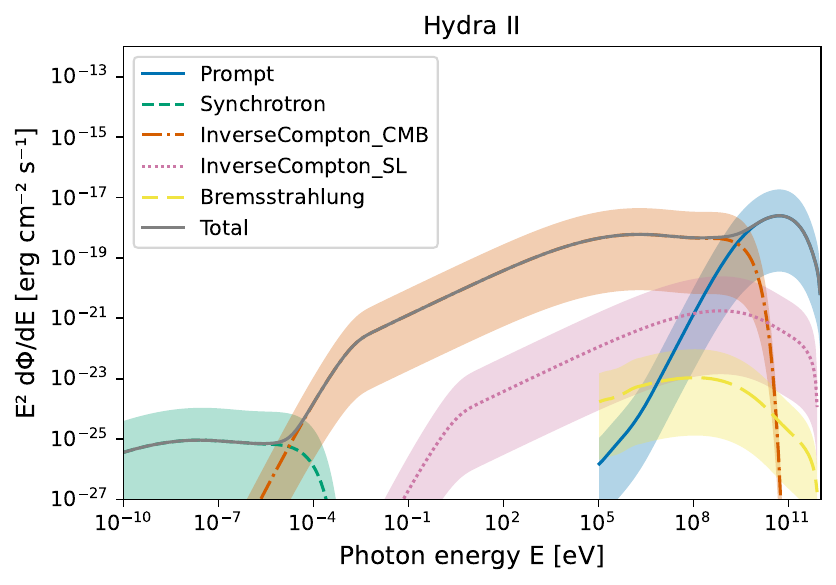}
		\caption{Same as Fig.\,\ref{fig:spec_Draco} but for selected dwarf spheroidals. \textbf{Upper left:} smallest distance / largest $J$-factor / largest total flux (Segue I). \textbf{Upper right:} Largest distance (Leo\,T). \textbf{Lower left:} Smallest $J$-factor (Leo\,IV). \textbf{Lower right:} Smallest total flux (Hydra\,II).} 
		\label{fig:spectra}
	\end{figure*}

\begin{figure}
		\centering
		\includegraphics[width=\columnwidth]{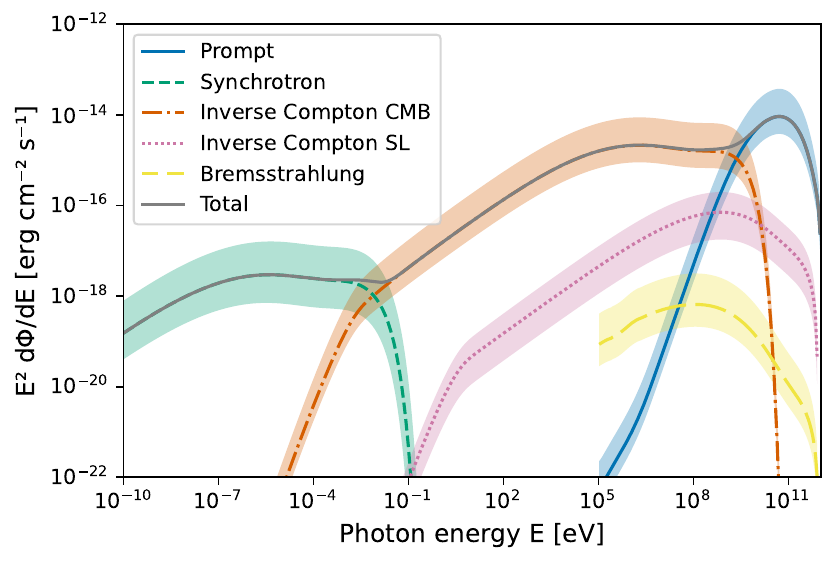}
		\caption{Same as Fig.\,\ref{fig:spec_Draco} but for the cumulative signal of all 27 selected dwarf galaxies.} 
		\label{fig:stacked}
	\end{figure}

	For each dSph the expected fluxes from DM annihilation are calculated as described in Sec.\,\ref{sec:multiwavelength}.
	In Fig.\,\ref{fig:spectra}, we show the modelled spectra for selected galaxies including uncertainties due to unknown properties such as magnetic field strengths, electron densities, and stellar population photon densities, as well as due to the calculations of the $J$-factors.
	Adding the fluxes of all dwarfs results in the total spectrum shown in Fig.\,\ref{fig:stacked}.
	
	%
	%Bands:
	
	%Erosita bands (0.5--2.0\,keV; 2.0--10.0\,keV)
	
	%Hard X-ray bands (NuSTAR, etc.: 10--20\,keV; 20--40\,keV; 40--80\,keV)
	
	%INTEGRAL bands (80--200\,keV; 200--500\,keV; 500-1000\,keV; 1--3\,MeV,3--10\,MeV, 10--30\,MeV)
	
	%Fermi/LAT and VHE bands (30--100\,MeV; 100-300\,MeV; 300-1000\,MeV; 1--3\,GeV; 3--10\,GeV; 10--30\,GeV; 30--100\,GeV; 100--300\,GeV; 300--1000\,GeV)
	
	We integrate the calculated fluxes over relevant instrument energy bands in the X-ray to the high-energy $\gamma$-ray range.
	The results are summarised in Tab.\,\ref{tab:expected_fluxes}, where the respective energy flux $F_{\text{E}}$ was determined by a weighted integration of the differential flux $d\Phi/dE$:
	\begin{equation}
		F_{\text{E}}=\int_{E_{\text{min}}}^{E_{\text{max}}} E \frac{d\Phi}{dE} dE\,\mathrm{,}
		\label{eq:energyflux}
	\end{equation}
	for which we use previous and current instrument bands to calculate the total flux within.
	In particular, we calculate integrated fluxes in the bands according to the sensitive range for INTEGRAL/ISGRI and JEM-X (1--30\,keV), Swift/BAT (14--195\,keV), INTEGRAL/SPI (0.1--1.0\,MeV), CGRO/COMPTEL (1--30\,MeV), and Fermi/LAT (30--100\,MeV and 0.1--100\,GeV), respectively.
	The total emission includes also the synchrotron emission and ranges from 0.1\,neV to 1\,TeV.

	In fact, also the neutrino emission from our selection of dwarf galaxies can be calculated using the prompt emission, Eq.\,(\ref{eq:flux}), with the neutrino spectra from Fig.\,\ref{fig:dNdE}.
	Here, one needs to take into account the oscillation of neutrinos which will change the fractions of \emph{measured} neutrinos at Earth.
	Individual galaxies would show a signal-to-background ratio in neutrinos around 50\,GeV of $10^{-6}$--$10^{-4}$ \citep{Spiering2012_CNB}, in which the background is mainly atmospheric at those energies.
	A calculation of the entire Cosmic Neutrino Background from this model is beyond the scope of this paper.

	\begin{table*}
		\begin{tabular}{|c|c|c|c|c|c|c|c|}
			\hline
			Galaxy & JEM-X|ISGRI & Swift/BAT & SPI & COMPTEL & Fermi/LAT & Fermi/LAT & Total \\
			& (1\,keV-30\,keV) & (14\,keV-195\,keV) & (100\,keV-1\,MeV) & (1\,MeV-30\,MeV) & (30\,MeV-100\,MeV) & (100\,MeV-100\,GeV) & (0.1\,neV-1\,TeV) \\
			\hline
			Carina & 2.05e-17 & 2.85e-17 & 3.35e-17 & 5.60e-17 & 1.52e-17 & 2.25e-16 & 4.38e-16 \\
			Draco & 1.59e-16 & 2.22e-16 & 2.60e-16 & 4.35e-16 & 1.18e-16 & 1.75e-15 & 3.40e-15 \\
			Fornax & 3.30e-17 & 4.60e-17 & 5.42e-17 & 9.21e-17 & 2.57e-17 & 3.94e-16 & 7.48e-16 \\
			Leo I & 1.01e-17 & 1.42e-17 & 1.68e-17 & 2.91e-17 & 8.45e-18 & 1.35e-16 & 2.48e-16 \\
			Leo II & 4.72e-18 & 6.58e-18 & 7.74e-18 & 1.30e-17 & 3.59e-18 & 5.42e-17 & 1.04e-16 \\
			Sculptor & 1.57e-17 & 2.44e-17 & 3.42e-17 & 1.05e-16 & 5.44e-17 & 1.21e-15 & 1.77e-15 \\
			Sextans & 2.12e-17 & 2.95e-17 & 3.47e-17 & 5.79e-17 & 1.56e-17 & 2.30e-16 & 4.49e-16 \\
			Ursa Minor & 2.87e-16 & 4.00e-16 & 4.71e-16 & 7.88e-16 & 2.14e-16 & 3.18e-15 & 6.17e-15 \\
			Boötes I & 8.46e-19 & 1.18e-18 & 1.38e-18 & 2.30e-18 & 6.21e-19 & 9.17e-18 & 1.79e-17 \\
			Coma Berenices & 8.45e-17 & 1.18e-16 & 1.38e-16 & 2.30e-16 & 6.20e-17 & 9.17e-16 & 1.79e-15 \\
			Canes Venatici I & 3.61e-18 & 5.02e-18 & 5.90e-18 & 9.83e-18 & 2.65e-18 & 3.91e-17 & 7.63e-17 \\
			Canes Venatici II & 8.25e-18 & 1.15e-17 & 1.35e-17 & 2.25e-17 & 6.06e-18 & 8.96e-17 & 1.75e-16 \\
			Hercules & 1.20e-18 & 1.66e-18 & 1.95e-18 & 3.26e-18 & 8.77e-19 & 1.30e-17 & 2.53e-17 \\
			Leo IV & 7.04e-19 & 9.80e-19 & 1.15e-18 & 1.92e-18 & 5.16e-19 & 7.63e-18 & 1.49e-17 \\
			Leo V & 1.28e-18 & 1.78e-18 & 2.09e-18 & 3.49e-18 & 9.39e-19 & 1.39e-17 & 2.71e-17 \\
			Leo T & 3.61e-18 & 5.03e-18 & 5.91e-18 & 9.90e-18 & 2.69e-18 & 4.01e-17 & 7.77e-17 \\
			Segue 1 & 4.72e-16 & 6.58e-16 & 7.73e-16 & 1.29e-15 & 3.47e-16 & 5.15e-15 & 1.01e-14 \\
			Segue 2 & 2.22e-18 & 3.09e-18 & 3.63e-18 & 6.06e-18 & 1.63e-18 & 2.41e-17 & 4.70e-17 \\
			Ursa Major I & 5.72e-17 & 7.97e-17 & 9.36e-17 & 1.56e-16 & 4.20e-17 & 6.20e-16 & 1.21e-15 \\
			Ursa Major II & 4.65e-16 & 6.47e-16 & 7.60e-16 & 1.27e-15 & 3.41e-16 & 5.04e-15 & 9.83e-15 \\
			Willman 1 & 3.76e-16 & 5.24e-16 & 6.16e-16 & 1.03e-15 & 2.77e-16 & 4.10e-15 & 7.98e-15 \\
			Reticulum II & 1.01e-16 & 1.41e-16 & 1.66e-16 & 2.77e-16 & 7.45e-17 & 1.10e-15 & 2.15e-15 \\
			Tucana II & 2.44e-16 & 3.40e-16 & 3.99e-16 & 6.65e-16 & 1.79e-16 & 2.64e-15 & 5.16e-15 \\
			Horologium I & 8.42e-17 & 1.17e-16 & 1.38e-16 & 2.30e-16 & 6.20e-17 & 9.17e-16 & 1.79e-15 \\
			Hydra II & 7.02e-19 & 9.77e-19 & 1.15e-18 & 1.91e-18 & 5.16e-19 & 7.63e-18 & 1.49e-17 \\
			Pisces II & 1.53e-17 & 2.13e-17 & 2.50e-17 & 4.18e-17 & 1.13e-17 & 1.67e-16 & 3.25e-16 \\
			Grus I & 1.77e-17 & 2.46e-17 & 2.89e-17 & 4.82e-17 & 1.30e-17 & 1.92e-16 & 3.74e-16 \\
			\hline
			Total & 2.49e-15 & 3.47e-15 & 4.09e-15 & 6.87e-15 & 1.88e-15 & 2.83e-14 & 5.45e-14 \\
			\hline
		\end{tabular}
		\caption{Expected fluxes from our scotogenic WIMP DM model including prompt emission from DM annihilation and secondary emission from inverse Compton scattering, bremsstrahlung, and synchrotron radiation, integrated over different energy bands. All values are calculated according to Eq.\,(\ref{eq:energyflux}) and given in units of erg\,cm$^{-2}$\,s$^{-1}$.}
		\label{tab:expected_fluxes}
	\end{table*}

	\section{Astrophysical fore- and background}\label{sec:astro_foreground}
	The expected primary and secondary emission of the scotogenic WIMP annihilation is subject to fore- and background emission along the lines-of-sight.
	For each galaxy, the Milky Way itself (diffuse emission and its own DM halo) as well as other high-energy sources, such as AGN, in small angular separations to the galaxies may contribute to the total signal.
	Background source confusion will become an issue only in the context of the angular resolutions of different high-energy telescopes as this changes roughly as function of energy as well as imaging technique (see Sect.\,\ref{sec:background_AGN}).
	We estimate all these contributions towards the directions of our selected galaxy sample to obtain a background-to-signal ratio in different bands for each galaxy.

	\subsection{Diffuse Galactic emission}\label{sec:diffuse_MW}
	We estimate the contribution of the diffuse emission in the Milky Way from hard X-rays to VHE photons by using GALPROP \citep{Strong2007_GALPROP} and the latest model by \citet{Bisschoff2019_Voyager1CR} that fits for the photon emission as well as the cosmic-ray distribution and abundances at Earth.
	It was shown by \citet{Siegert2022_MWdiffuse} that this model underestimates the diffuse emission in the MeV band and that either a slightly changed diffusion model or enhanced optical interstellar radiation field is required.
	For the sake of consistency with the literature models, we focus on the much more broadly tested model by \citet{Bisschoff2019_Voyager1CR}.
	We use GALPROP version 57 \citep{Porter2022_GALPROPv57} to estimate the contributions of the diffuse emission above 100\,keV including bremsstrahlung, pion decay, and Inverse Compton scattering.

\begin{figure*}
		\centering
		
		\includegraphics[width=0.7\columnwidth]{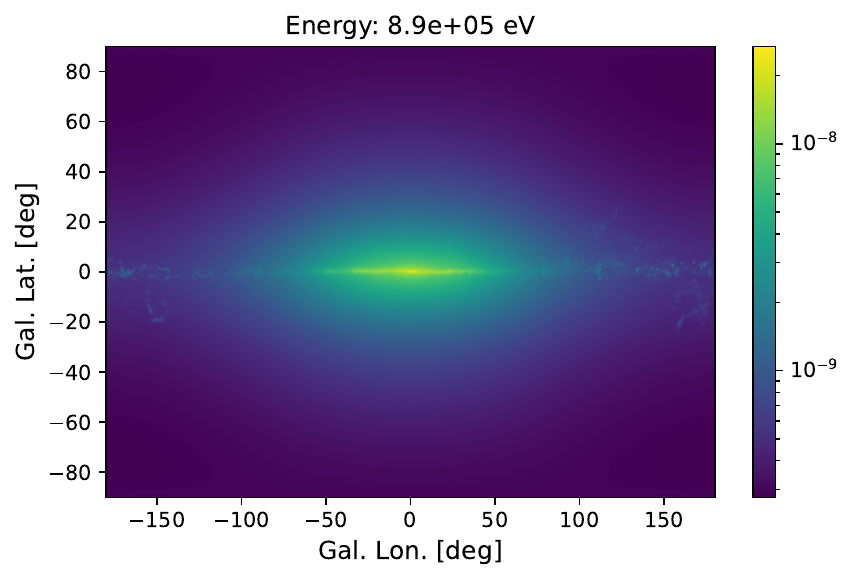}~
		\includegraphics[width=0.7\columnwidth]{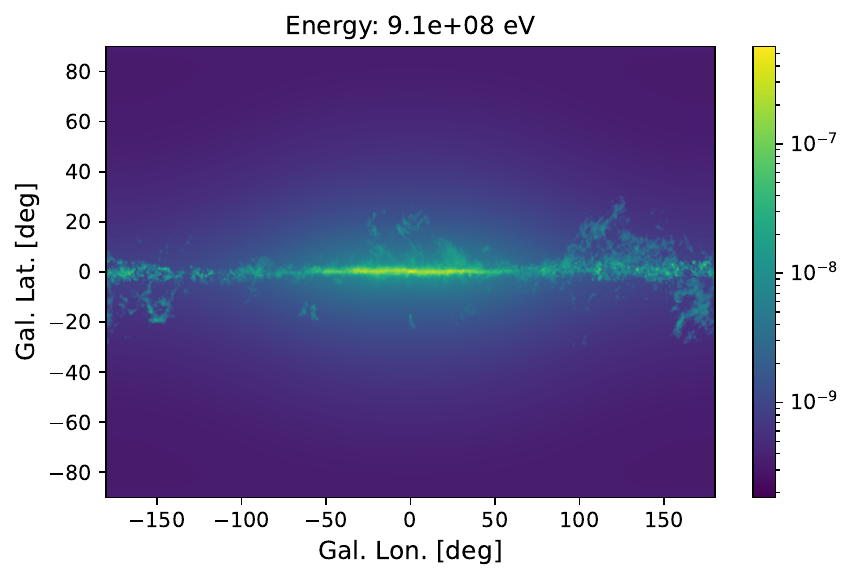}~
		\includegraphics[width=0.7\columnwidth]{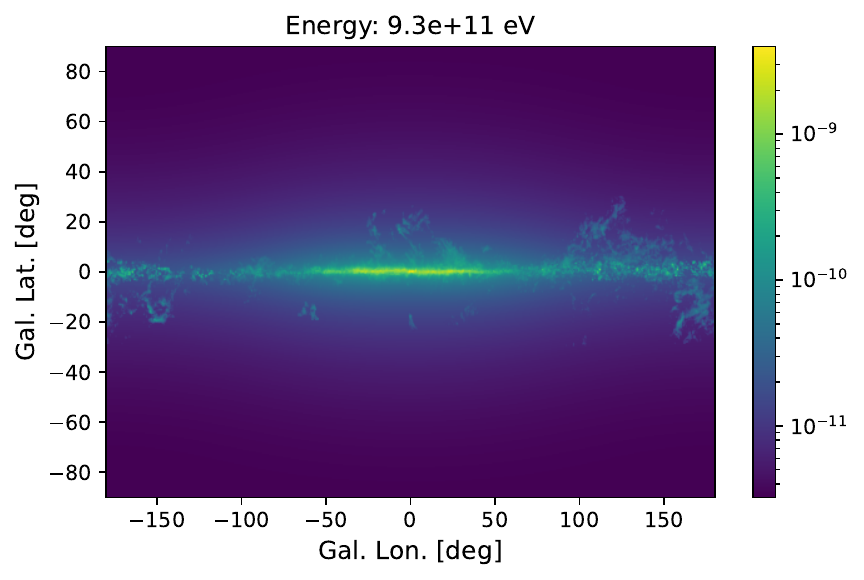}
		
		\caption{GALPROP (v57) simulations at different energies (left: 1\,MeV; middle: 1\,GeV; right: 1\,TeV) including fluxes from bremsstrahlung, Inverse Compton scattering and pion decay. Fluxes are given in units of $\mathrm{erg\,cm^{-2}\,s^{-1}\,sr^{-1}}$.} 
		\label{fig:galprop_maps}
	\end{figure*}

\begin{figure*}
		\centering
		
		\includegraphics[width=0.7\columnwidth]{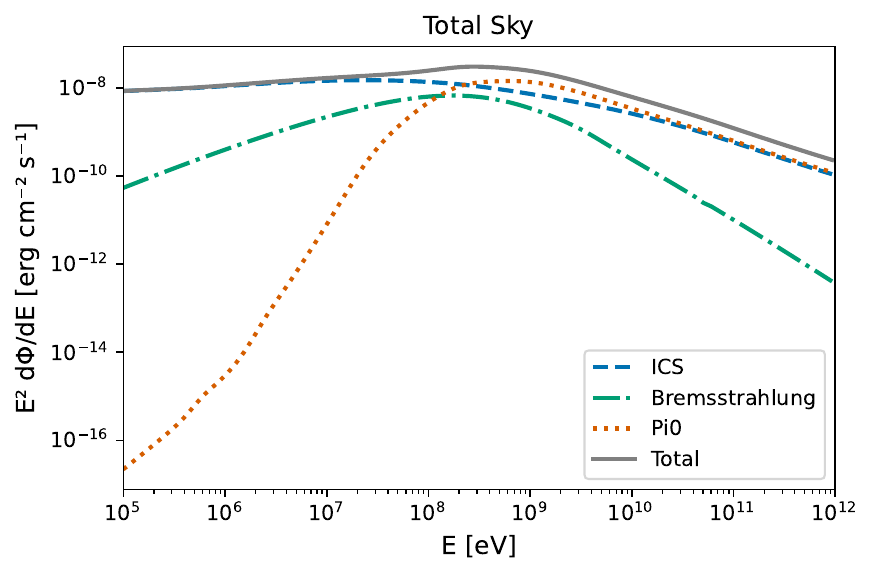}~
		\includegraphics[width=0.7\columnwidth]{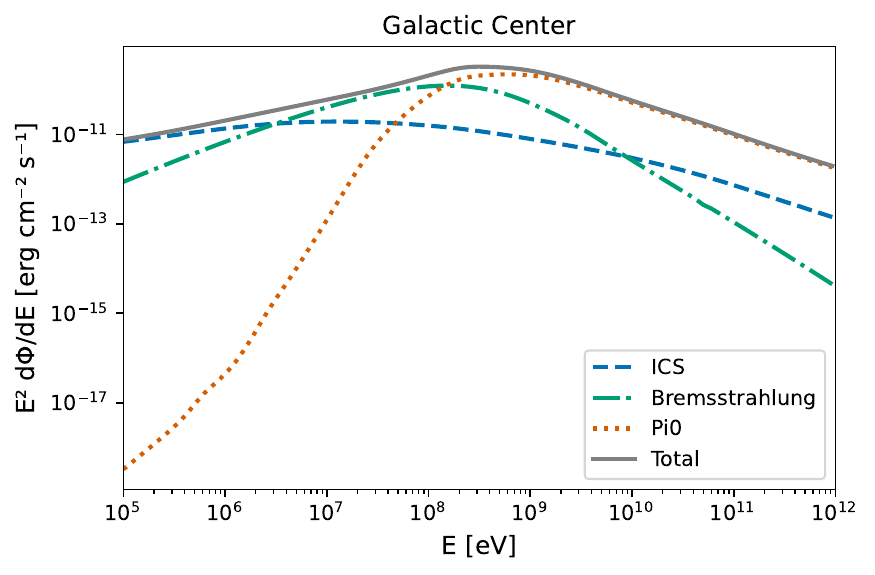}~
		\includegraphics[width=0.7\columnwidth]{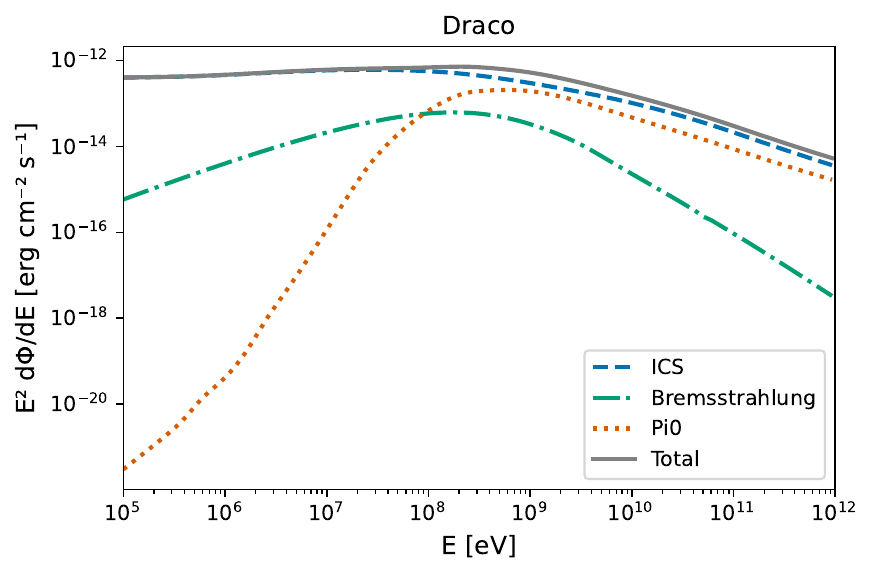}
		
		\caption{GALPROP (v57) spectra including Inverse Compton scattering, bremsstrahlung, and pion decay of the total sky (left), towards the direction of the Galactic Center (middle), and towards the direction of Draco (right). The latter two were calculated for an angular region of $1^\circ$ each.} 
		\label{fig:galprop_spectra}
	\end{figure*}

	In Fig.\,\ref{fig:galprop_maps}, we show examples of how the diffuse Milky Way foreground appears as a function of photon energy.
	Similarly, in Fig.\,\ref{fig:galprop_spectra}, we show the resulting spectra of diffuse emission for the total sky, the Galactic Center ($1^\circ$ region), and towards the direction of Draco ($1^\circ$ region).
	Around 1\,MeV, the total diffuse emission is completely dominated by Inverse Compton scattering ($>99$\%) with only small contributions from bremsstrahlung.
	In the image, this is reflected in the largely missing structure from the gas in the Galaxy, being outshined by the Inverse Compton glow to high latitudes.
	Above 1\,GeV, the gas structure of the Milky Way is clearly visible as cosmic-ray induced pion production is showing directly where the gas of the Galaxy is found.
	At 1\,TeV the Inverse Compton and pion decay structure are equally bright.
	In the spectrum, around 0.5\,GeV, the flux levels of bremsstrahlung, Inverse Compton and pion decay are comparable and pion decay takes over as strongest emission component from 1\,GeV up to 1\,TeV.
	In the Galactic Center, the emission below 5\,MeV is mainly Inverse Compton, from 5--100\,MeV is dominated by bremsstrahlung, and above by pion decay.
	In the case of Draco, being located at Galactic coordinates $\ell = 86.37^\circ$, $b = 34.72^\circ$, the presence of gas is minimal and only the Inverse Compton component is significant up to $\sim 1$\,GeV.
	Above, a $\sim 10$\% fraction of the flux can also be expected from pion decay.
	
	As a summary in Tab.\,\ref{tab:galprop_fluxes}, we show the diffuse Galactic emission fluxes integrated across different bands as typically measured by different instruments.
	The diffuse emission of the Galaxy in hard X-rays, neglecting the strong contribution from unresolved point sources below $\sim 100$\,keV \citep{Krivonos2007_GRXE}, is very uncertain given the current measurements.
	For this reason, we only quote the simulation results above 100\,keV in Tab.\,\ref{tab:galprop_fluxes}.
	Considering the works by \citet{Krivonos2007_GRXE,Bouchet2011_diffuseCR} and \citet{Siegert2022_diffuseemission}, the inner Galaxy flux around 100\,keV, including unresolved and resolved point sources, is already $10^{-8}\,\mathrm{erg\,cm^{-2}\,s^{-1}}$, so that the total Galactic contribution may be even higher than the left panel of Fig.\,\ref{fig:galprop_spectra}.
	Extrapolating down to 10\,keV, the total diffuse flux may be as high as $10^{-7}\,\mathrm{erg\,cm^{-2}\,s^{-1}}$ for the entire sky.
	Likewise, between 200 and 511\,keV, the additional component of positron annihilation in the form of Positronium (Ps) decay (ortho-Ps continuum and para-Ps line) would enhance the diffuse emission in the Galactic bulge and disk \citep{Siegert2016_511,Siegert2023_511}.
	In fact, the Ps decay makes the largest contribution in terms of foreground emission in the SPI band ($0.1$--$1.0$\,MeV), amounting to $\sim 9 \times 10^{-9}\,\mathrm{erg\,cm^{-2}\,s^{-1}}$ for the full sky, and about $3 \times 10^{-10}\,\mathrm{erg\,cm^{-2}\,s^{-1}}$ towards the Galactic Center.
	In the direction of Draco, the Galactic Ps contribution is weak and we estimate $4 \times 10^{-16}\,\mathrm{erg\,cm^{-2}\,s^{-1}}$ based on the model by \citet{Siegert2016_511}.
	An inclusion of the tertiary component from positron annihilation of DM annihilation byproducts is beyond the scope of this work and will be addressed in a future study.

	\begin{table*}
		\begin{tabular}{|c|c|c|c|c|c|c|}
			\hline
			Galaxy & JEM-X|ISGRI & Swift/BAT & SPI & COMPTEL & Fermi/LAT & Fermi/LAT \\
			& (1\,keV-30\,keV) & (14\,keV-195\,keV) & (100\,keV-1\,MeV) & (1\,MeV-30\,MeV) & (30\,MeV-100\,MeV) & (100\,MeV-100\,GeV) \\
			\hline
			Carina & - & - & 1.89e-11 & 2.04e-11 & 6.71e-12 & 3.22e-12 \\
			Draco & - & - & 1.53e-11 & 1.73e-11 & 5.65e-12 & 2.44e-12 \\
			Fornax & - & - & 9.15e-12 & 1.03e-11 & 3.35e-12 & 1.48e-12 \\
			Leo I & - & - & 1.01e-11 & 1.10e-11 & 3.60e-12 & 1.60e-12 \\
			Leo II & - & - & 8.93e-12 & 1.00e-11 & 3.27e-12 & 1.48e-12 \\
			Sculptor & - & - & 9.09e-12 & 1.05e-11 & 3.40e-12 & 1.47e-12 \\
			Sextans & - & - & 1.16e-11 & 1.26e-11 & 4.14e-12 & 1.83e-12 \\
			Ursa Minor & - & - & 1.16e-11 & 1.31e-11 & 4.28e-12 & 1.85e-12 \\
			Boötes I & - & - & 1.05e-11 & 1.25e-11 & 4.04e-12 & 1.65e-12 \\
			Coma Berenices & - & - & 8.86e-12 & 1.01e-11 & 3.29e-12 & 1.42e-12 \\
			Canes Venatici I & - & - & 9.22e-12 & 1.06e-11 & 3.46e-12 & 1.50e-12 \\
			Canes Venatici II & - & - & 8.83e-12 & 9.93e-12 & 3.24e-12 & 1.42e-12 \\
			Hercules & - & - & 2.73e-11 & 3.58e-11 & 1.12e-11 & 4.39e-12 \\
			Leo IV & - & - & 1.05e-11 & 1.19e-11 & 3.89e-12 & 1.74e-12 \\
			Leo V & - & - & 1.02e-11 & 1.15e-11 & 3.75e-12 & 1.68e-12 \\
			Leo T & - & - & 1.04e-11 & 1.12e-11 & 3.67e-12 & 1.70e-12 \\
			Segue 1 & - & - & 9.85e-12 & 1.09e-11 & 3.56e-12 & 1.62e-12 \\
			Segue 2 & - & - & 1.09e-11 & 1.16e-11 & 3.82e-12 & 1.78e-12 \\
			Ursa Major I & - & - & 9.29e-12 & 1.01e-11 & 3.32e-12 & 1.50e-12 \\
			Ursa Major II & - & - & 1.10e-11 & 1.17e-11 & 3.86e-12 & 1.86e-12 \\
			Willman 1 & - & - & 9.19e-12 & 1.01e-11 & 3.31e-12 & 1.54e-12 \\
			Reticulum II & - & - & 1.13e-11 & 1.29e-11 & 4.21e-12 & 1.90e-12 \\
			Tucana II & - & - & 1.35e-11 & 1.64e-11 & 5.27e-12 & 2.16e-12 \\
			Horologium I & - & - & 1.08e-11 & 1.25e-11 & 4.06e-12 & 1.79e-12 \\
			Hydra II & - & - & 2.01e-11 & 2.36e-11 & 7.63e-12 & 3.23e-12 \\
			Pisces II & - & - & 1.25e-11 & 1.45e-11 & 4.70e-12 & 2.02e-12 \\
			Grus I & - & - & 1.23e-11 & 1.49e-11 & 4.80e-12 & 2.00e-12 \\
			\hline
		\end{tabular}
		\caption{Diffuse Galactic emission foreground of the Milky Way, calculated with GALPROP (v57) in different energy bands in units of erg\,cm$^{-2}$\,s$^{-1}$. Because the diffuse emission below 100\,keV is much enhanced by unresolved point sources, we only quote values for 100\,keV and above.}
		\label{tab:galprop_fluxes}
	\end{table*}

	\subsection{Dark matter halo of the Galaxy}\label{sec:MW_DM_halo}
	We estimate the prompt emission from the scotogentic WIMP model in the Milky Way itself by line-of-sight-integrating a standard NFW profile \citep{Navarro1997_NFW}.
	The resulting fluxes for the directions of our 27 selected galaxies are listed in Tab.\,\ref{tab:NFW_fluxes}.
	Also here, the fluxes below 100\,keV are omitted, however now because the fluxes are small from the prompt emission alone.
	The quoted fluxes directly reflect the locations of the dwarf galaxies which show larger fluxes the closer they are to the Galactic Center.
	The peak emission is again found around 50\,GeV and the total flux is about five orders of magnitude below the expected diffuse emission.
	Likewise, for the direction of the Galactic Center, the Galactic foreground is outshining the prompt scotogenic DM annihilation emission by two to three orders of magnitude.
	This means on the basis of prompt emission alone, there is no chance to detecting the prompt emission with this or even the next generation of high-energy telescopes.
	What is required is the broadband spectrum of the primary, secondary, and potentially tertiary component of DM annihilation products and the cumulative effect of different targets to avoid the strong Galactic foreground.

	The secondary emission of transported DM annihilation products is not trivial.
	\citet{Cirelli2009_IC_darkmatter,Cirelli2011_DM,Cirelli2021_subGeVDM,Cirelli2023_subGeVDM}, for example, showed that one can construct a `halo function' of charged particles in the Galaxy experiencing the conditions of the interstellar medium and interstellar radiation field.
	This yields to a specific shape of the Inverse Compton halo in the MeV band, for example, as well as shapes for the other processes.
	In this work, we omit the distribution of the secondary emission from the Milky Way DM halo itself because large parts of what could be expected from DM annihilation is already taken into account by the GALPROP estimates for diffuse emission.
	The model by \citet{Bisschoff2019_Voyager1CR}, for example, already tries to explain the \emph{entire} diffuse emission and cosmic-ray abundances, so that using the emission of charged particles from DM annihilation in the Milky Way would `double count' possible foreground.
	We note, however, that these contributions are important and should be carefully taken into account when multiwavelength measurements are attempted.

	\begin{table*}
		\begin{tabular}{|c|c|c|c|c|c|c|c|}
			\hline
			Galaxy & JEM-X|ISGRI & Swift/BAT & SPI & COMPTEL & Fermi/LAT & Fermi/LAT & Total \\
			& (1\,keV-30\,keV) & (14\,keV-195\,keV) & (100\,keV-1\,MeV) & (1\,MeV-30\,MeV) & (30\,MeV-100\,MeV) & (100\,MeV-100\,GeV) & (100\,keV-1\,TeV) \\
			\hline
			Carina & - & - & 2.95e-22 & 6.25e-20 & 5.86e-19 & 8.92e-16 & 1.59e-14 \\
			Draco & - & - & 3.66e-22 & 8.13e-20 & 7.63e-19 & 9.93e-16 & 1.97e-14 \\
			Fornax & - & - & 2.87e-22 & 5.90e-20 & 5.54e-19 & 5.56e-16 & 1.54e-14 \\
			Leo I & - & - & 2.38e-22 & 5.02e-20 & 4.71e-19 & 4.89e-16 & 1.28e-14 \\
			Leo II & - & - & 2.69e-22 & 5.70e-20 & 5.35e-19 & 7.46e-16 & 1.44e-14 \\
			Sculptor & - & - & 3.76e-22 & 7.88e-20 & 7.39e-19 & 1.07e-15 & 2.02e-14 \\
			Sextans & - & - & 2.59e-22 & 5.44e-20 & 5.10e-19 & 6.11e-16 & 1.39e-14 \\
			Ursa Minor & - & - & 2.97e-22 & 6.56e-20 & 6.16e-19 & 6.67e-16 & 1.59e-14 \\
			Boötes I & - & - & 5.78e-22 & 1.23e-19 & 1.16e-18 & 1.27e-15 & 3.11e-14 \\
			Coma Berenices & - & - & 3.34e-22 & 6.98e-20 & 6.55e-19 & 7.83e-16 & 1.79e-14 \\
			Canes Venatici I & - & - & 3.82e-22 & 8.03e-20 & 7.53e-19 & 1.05e-15 & 2.05e-14 \\
			Canes Venatici II & - & - & 2.82e-22 & 5.99e-20 & 5.62e-19 & 6.12e-16 & 1.52e-14 \\
			Hercules & - & - & 2.05e-21 & 4.25e-19 & 3.99e-18 & 5.67e-15 & 1.10e-13 \\
			Leo IV & - & - & 3.33e-22 & 6.85e-20 & 6.42e-19 & 5.97e-16 & 1.79e-14 \\
			Leo V & - & - & 3.25e-22 & 6.66e-20 & 6.24e-19 & 6.85e-16 & 1.75e-14 \\
			Leo T & - & - & 2.06e-22 & 4.43e-20 & 4.16e-19 & 6.06e-16 & 1.11e-14 \\
			Segue 1 & - & - & 2.38e-22 & 4.67e-20 & 4.38e-19 & 4.66e-16 & 1.28e-14 \\
			Segue 2 & - & - & 2.02e-22 & 4.19e-20 & 3.93e-19 & 6.23e-16 & 1.09e-14 \\
			Ursa Major I & - & - & 2.10e-22 & 4.37e-20 & 4.10e-19 & 4.06e-16 & 1.13e-14 \\
			Ursa Major II & - & - & 1.91e-22 & 3.70e-20 & 3.47e-19 & 4.95e-16 & 1.03e-14 \\
			Willman 1 & - & - & 2.22e-22 & 4.75e-20 & 4.46e-19 & 5.86e-16 & 1.19e-14 \\
			Reticulum II & - & - & 3.36e-22 & 7.19e-20 & 6.74e-19 & 8.74e-16 & 1.81e-14 \\
			Tucana II & - & - & 8.00e-22 & 1.76e-19 & 1.65e-18 & 2.35e-15 & 4.30e-14 \\
			Horologium I & - & - & 3.65e-22 & 7.77e-20 & 7.29e-19 & 9.95e-16 & 1.96e-14 \\
			Hydra II & - & - & 5.81e-22 & 1.24e-19 & 1.16e-18 & 1.65e-15 & 3.12e-14 \\
			Pisces II & - & - & 4.28e-22 & 8.84e-20 & 8.30e-19 & 1.13e-15 & 2.30e-14 \\
			Grus I & - & - & 7.56e-22 & 1.62e-19 & 1.52e-18 & 2.22e-15 & 4.06e-14 \\
			\hline
			total & - & - & 1.12e-20 & 2.36e-18 & 2.22e-17 & 2.91e-14 & 6.02e-13 \\
			\hline
		\end{tabular}
		\caption{Contributions of the prompt scotogenic WIMP dark matter annihilation in the Milky Way assuming an NFW-profile as dark matter halo. The fluxes are given in units of erg\,cm$^{-2}$\,s$^{-1}$.}
		\label{tab:NFW_fluxes}
	\end{table*}

	\subsection{Source confusion}\label{sec:background_AGN}
	Given that different high-energy instruments have different angular resolutions at different energies, we consider source catalogues of the respective instruments to search for nearby sources which would enhance the fore- and background emission compared to DM-only signal.
	Possible background emission comes mostly from AGN in the vicinity of our selected galaxies.
	In order to identify such sources in the direction of the respective dwarf galaxy, we consider the following catalogues covering the whole energy spectrum from X-rays to high-energies:

	The INTEGRAL General Reference Source Catalogue \citep{Bird2010_ibis_cat} with sources detected by INTEGRAL/ISGRI (20--200\,keV), the Swift/BAT 105-Month Hard X-ray Catalogue \citep{swiftcatalog} covering 14\,keV-195\,keV, the SPI Catalogue covering 20\,keV--2\,MeV \citep{Bouchet2008_imaging}, the First COMPTEL Source Catalogue \citep{comptel} covering 1\,MeV--30\,MeV, the First Fermi-LAT Low Energy Catalogue \citep{1FLE} covering 30\,MeV-100\,MeV, and the Fermi-LAT Fourth Source Catalogue covering 100\,MeV-100\,GeV.
	%
	%\LE{
		%\begin{itemize}
		%    \item 1\,keV-30\,keV: the INTEGRAL General Reference Source Catalogue \citep{website?}
		%    \item 14\,keV-195\,keV: the Swift/BAT 105-Month Hard X-ray Catalogue \citep{swiftcatalog} 
		%    \item 100\,keV-1\,MeV: the SPI catalogue
		%    \item 1\,MeV-30\,MeV: the First COMPTEL Source Catalogue \citep{comptel}
		%    \item 30\,MeV-100\,MeV: the First Fermi-LAT Low Energy Catalogue \citep{1FLE}
		%    \item 100\,MeV-100\,GeV: the Fermi-LAT Fourth Source Catalogue \citep{Abdollahi_2020}
		%\end{itemize}
		%}
	%
	We calculate the angular distance of all dwarf galaxies to all catalogue sources and consider an angular separation of $0.6^\circ$, $1^\circ$, $4^\circ$, $3^\circ$, and $3^\circ$ and $1^\circ$ for ISGRI, Swift/BAT, SPI, COMPTEL, and Fermi/LAT, respectively, as potentially contributing considering the halo sizes of the dwarfs to be most less than $1^\circ$.
	The assumed angular separation in each energy band, depending on the instruments' resolutions, and potential background sources as well as their energy fluxes are displayed in Tab.\,\ref{tab:background_AGN}.
	In total, 15 AGN coincide with 11 of the 27 dwarf galaxies in the selected angular ranges.
	While the same sources might also contribute to the radio bands, we omit the cross correlations with radio sources because within one square-degree, there are dozens if not hundreds of sources.
	
	It is clear that these sources may lead to source confusion if the dwarf galaxies are analysed individually.
	Especially in the wings of the point spread functions, some addition contribution may be neglected if the signal is particularly weak.
	For this reason, the galaxies would be analysed in a coherent manner as they all share the underlying model.
	This means that only one (or two) dark matter parameters are shared by the dwarf galaxies which makes any DM annihilation model more rigid and therefore more predictive.

	\begin{table*}
		\begin{tabular}{|c|c|c|c|c|c|c|}
			\hline
			& JEM-X|ISGRI (0.6°) & Swift/BAT (1°) & SPI (4°) & COMPTEL (3°) & Fermi/LAT (3°) & Fermi/LAT (1°) \\
			& (1\,keV-30\,keV) & (14\,keV-195\,keV) & (100\,keV-1\,MeV) & (1\,MeV-30\,MeV) & (30\,MeV-100\,MeV) & (100\,MeV-100\,GeV) \\
			\hline
			Carina & - & SWIFT J0639.9-5124 & - & - & - & - \\
			&  & 7.92e-12 &  &  &  &  \\
			%  &  &  &  &  &  \\
			Fornax & - & - & - & - & - & 4FGL J0240.8-3401 \\
			&  &  &  &  &  & 1.95e-12 \\
			%  &  &  &  &  &  \\
			Sextans & - & - & - & - & - & 4FGL J1010.8-0158 \\
			&  &  &  &  &  & 4.61e-12 \\
			%  &  &  &  &  &  \\
			Coma Berenices & - & - & - & GRO J1224+2155 & 1FLE J1224+2118 & 4FGL J1224.4+2436 \\
			&  &  &  & 2.71e-10 & 5.64e-11 & 1.61e-11 \\
			%  &  &  &  &  &  \\
			Canes Venatici II & - & - & NGC 4138 & - & - & - \\
			&  &  & 2.45e-08 &  &  &  \\
			%  &  &  &  &  &  \\
			Hercules & - & SWIFT J1650.5+0434 & - & - & - & 4FGL J1650.9+0429 \\
			&  & 2.08e-11 &  &  &  & 2.36e-12 \\
			&  &  &  &  &  & 4FGL J1649.6+0411 \\
			&  &  &  &  &  & 3.37e-12 \\
			%  &  &  &  &  &  \\
			Willman 1 & - & - & - & - & - & 4FGL J1049.7+5011 \\
			&  &  &  &  &  & 2.11e-12 \\
			%  &  &  &  &  &  \\
			Tucana II & - & - & - & - & - & 4FGL J2247.7-5857 \\
			&  &  &  &  &  & 2.08e-12 \\
			%  &  &  &  &  &  \\
			Horologium I & - & - & - & - & - & 4FGL J0253.2-5441 \\
			&  &  &  &  &  & 4.23e-12 \\
			%  &  &  &  &  &  \\
			Pisces II & - & SWIFT J2256.5+0526 & - & - & - & - \\
			&  & 9.11e-12 &  &  &  &  \\
			%  &  &  &  &  &  \\
			Grus I & J225400.0-500000 & - & - & - & - & - \\
			& 2.55e-21 &  &  &  &  &  \\
			\hline
		\end{tabular}
		\caption{Background AGN in the vicinity of our selected dwarf galaxies. The fluxes are given in units of erg\,cm$^{-2}$\,s$^{-1}$. The assumed angular separation between the dwarf galaxies and the sources of the selected catalogues (see Sec.\,\ref{sec:background_AGN}) is indicated in parentheses after the instrument in each energy band. In total, 15 AGN coincide with one of the dwarf galaxies in the selected angular ranges.}
		\label{tab:background_AGN}
	\end{table*}

	\subsection{Signal to background ratio}\label{sec:SNB}
	When comparing the individual fluxes of the selected samples of dwarf galaxies to the possible fore- and background emission components from diffuse emission and nearby point sources, it is evident that no galaxy would outshine the Milky Way.
	The expected double- (X-rays to VHE $\gamma$-rays) or triple-hump (including radio) structure of the DM emission in the dwarf galaxies is very characteristic and unlike any other emission mechanism.
	This makes it particularly suited for multi-wavelength analyses including as many high-energy (survey) instruments as possible.
	While the expected emission in either of the considered energy bands is between two and three orders of magnitude below the fore- and background contributions, long exposures and multiple sources with a common amplitude parameter will help to disentangle the emission components.

	To enhance the signal-to-background ratio even further, more (nearby) galaxies should be taken into account, such as the catalogue of (now) more than 1500 Local Volume Galaxies \citep{Karachentsev2006_galaxydistances} within a distance of 11\,Mpc.
	The supergalactic plane will then create an apparent structure distinct from any expected X- and $\gamma$-ray signal.
	Likewise, the tertiary emission from position annihilation would also provide much more leverage unless the positrons escape into the intergalactic medium.

	\section{Discussion and conclusion}\label{sec:discussion}
	In this work, we studied the prompt and secondary emission from scotogenic WIMP annihilations in 27 dwarf galaxies and compared the fluxes to diffuse and pointlike back- and foreground sources.
	We find that in the X- to $\gamma$-ray bands from 1\,keV to 1\,TeV, the emission level of individual galaxies is between $10^{-19}$--$10^{-15}\,\mathrm{erg\,cm^{-2}\,s^{-1}}$, with a total flux of all galaxies amounting to $10^{-15}$--$10^{-13}\,\mathrm{erg\,cm^{-2}\,s^{-1}}$.
	This is in stark contrast to the diffuse emission at these energies towards the directions of the satellite galaxies, on the order of $10^{-12}$--$10^{-11}\,\mathrm{erg\,cm^{-2}\,s^{-1}}$.
	The signal-to-background ratio for individual galaxies is therefore in the range of $10^{-3}$--$10^{-2}$.
	In order to `outshine' the diffuse emission, the galaxies must therefore be exposed for very long times.
	However, the galaxies should not be analysed one-by-one with single instruments because they share the common parameters from the scotogenic WIMP and therefore should be treated as `one entity' instead of 27.
	Certainly, such an analysis is similar to `stacking' but which is difficult across different instruments.
	Instead a Bayesian Hierarchical Model should be employed, so that also the uncertainties from the galaxies themselves, such as their interstellar radiation field, magnetic field, and electron densities, can properly be taken into account.
	In such a framework, also more targets can be embedded if their intrinsic emission can be properly modelled and if enough data are available.
	Building a DM annihilation model for the more than 1500 Local Volume Galaxies, together with their intrinsic emission and possible foreground from the Milky Way, will be left for a future study.

	While the Milky Way itself would certainly show the strongest signal -- which is true for most DM models --, the strong Galactic foreground prevents clear and robust estimates of DM properties unless all known components are taken into account with their respective uncertainties.
	Especially in the MeV band where the second `hump' from the (scotogenic) WIMP secondary emission is found, the diffuse emission is hardly understood \citep[e.g.,][]{Siegert2022_MWdiffuse}.
	Therefore, even though the Milky Way would share the same prompt emission which can easily be modelled, the secondary and tertiary emission in the Milky Way requires a more sophisticated treatment \citep[e.g.,][]{Cirelli2009_IC_darkmatter,Cirelli2011_DM,Cirelli2021_subGeVDM,Cirelli2023_subGeVDM}.
	Analysing other galaxies -- together with the Milky Way -- may provide more stringent constraints on DM particle properties if analysed coherently for prompt, secondary, and tertiary emission.

	Even though individual galaxies are found to be weak in scotogenic DM annihilation (primary and secondary), some sources may stick out earlier than others.
	Typically, and this is also the case for our work, these are the galaxies Ursa Major II (1--30\,keV, 1--30\,MeV) and Segue I (14--1000\,keV, 0.03--100\,GeV). 
	In terms of total flux, the list can be extended to Ursa Major I, Ursa Minor, Willman 1, Reticulum II, Tucana II, Horologium I, Coma Berenices, Scultor, and Draco, all showing a total flux above $10^{-15}\,\mathrm{erg\,cm^{-2}\,s^{-1}}$.
	We note that these estimates only depend weakly on the galaxy parameters because their sizes are well constrained and the local CMB photon density is known.
	The total emission uncertainty is mainly due to the $J$-factors with an average uncertain of 50\%.

	Considering future measurements, wide field-of-few survey instruments are probably the best choice for studying (scotogenic) DM annihilation emission from X- to $\gamma$-rays.
	Certainly also radio telescope arrays, such as NenuFAR \citep{nenufar}, with a coarse angular resolution would help in disentangling the DM emission from Galactic backgrounds.
	Considering GeV--TeV emission, individual sources may obtain a large accuracy boost from the full Cerenkov Telescope Array (CTA), which, in combination with Fermi/LAT, will provide deep exposures.
	In order to observe hundreds of sources for long times, the LAT is particularly indispensable.
	In particular in the MeV range, the Compton Spectrometer and Imager \citep[COSI;][]{Tomsick2019_COSI,Tomsick2023_COSI} will launch in 2027 and enhance the sensitivity from 0.2--5\,MeV immensely compared to previous instruments.
	INTEGRAL/SPI will not be operating beyond 2024 so that its archival data can be used to constrain individual sources if they had been observed.
	Most of the exposure for INTEGRAL had been spent on the Galactic bulge and plane, and it only rarely observed higher latitudes.
	COSI, on the other hand, will survey the full sky once per day with a field-of-view of about 25\% of the sky.
	Together with Fermi/LAT, COSI will enable systematic searches for multiple DM targets, and in particular be sensitive to the emission peak in the MeV band from secondary (Inverse Compton, bremsstrahlung) and tertiary (positron annihilation) interactions of DM annihilation products.

	Since the individual sources show a signal-to-background ratio of $10^{-3}$--$10^{-2}$, an important way to disentangling (scotogenic) DM signals from foreground is found in studies of the Cosmic Gamma-ray Background (CGB).
	The secondary and tertiary components would show a very peculiar spectrum as the red-shift evolution of the CMB as well as the dependence of the resulting spectrum integrated along red-shifts has to be taken into account.
	In the case of the secondary MeV bump from Inverse Compton scattering, on the one hand, the CGB feature would be broadened;
	for the tertiary emission of the 511\,keV line, on the other hand, only a one-sided shift and broadening would be expected \citep{Iguaz2021_CGBPBH}.
	In addition would the angular power spectrum of primary, secondary, and tertiary components of DM in the CGB show distinct features compared to the AGN contribution, for example.
	With COSI, these measurements become possible for the first time, opening a new window for the indirect search of DM.

	% Example equation
	%\begin{equation}
	%    x=\frac{-b\pm\sqrt{b^2-4ac}}{2a}.
	%	\label{eq:quadratic}
	%\end{equation}
	
	%\begin{figure}
	%	% To include a figure from a file named example.*
	%	% Allowable file formats are eps or ps if compiling using latex
	%	% or pdf, png, jpg if compiling using pdflatex
	%	\includegraphics[width=\columnwidth]{example}
	%    \caption{This is an example figure. Captions appear below each figure.
		%	Give enough detail for the reader to understand what they're looking at,
		%	but leave detailed discussion to the main body of the text.}
	%    \label{fig:example_figure}
	%\end{figure}
	
	%\begin{table}
	%	\centering
	%	\caption{This is an example table. Captions appear above each table.
		%	Remember to define the quantities, symbols and units used.}
	%	\label{tab:example_table}
	%	\begin{tabular}{lccr} % four columns, alignment for each
		%		\hline
		%		A & B & C & D\\
		%		\hline
		%		1 & 2 & 3 & 4\\
		%		2 & 4 & 6 & 8\\
		%		3 & 5 & 7 & 9\\
		%		\hline
		%	\end{tabular}
	%\end{table}

	%%%%%%%%%%%%%%%%%%%%%%%%%%%%%%%%%%%%%%%%%%%%%%%%%%
	\section*{Data Availability}
	The emission models are available as \texttt{astromodels}\footnote{\url{https://astromodels.readthedocs.io/en/latest/index.html}} template functions upon reasonable request.
	
	%The inclusion of a Data Availability Statement is a requirement for articles published in MNRAS. Data Availability Statements provide a standardised format for readers to understand the availability of data underlying the research results described in the article. The statement may refer to original data generated in the course of the study or to third-party data analysed in the article. The statement should describe and provide means of access, where possible, by linking to the data or providing the required accession numbers for the relevant databases or DOIs.

	
	
	\bibliographystyle{mnras}
	\bibliography{example,thomas,etn} % if your bibtex file is called example.bib

	%\begin{thebibliography}{99}
	%\end{thebibliography}
	
	%%%%%%%%%%%%%%%%%%%%%%%%%%%%%%%%%%%%%%%%%%%%%%%%%%
	
	%%%%%%%%%%%%%%%%% APPENDICES %%%%%%%%%%%%%%%%%%%%%
	
	%\appendix
	
	
	
	%%%%%%%%%%%%%%%%%%%%%%%%%%%%%%%%%%%%%%%%%%%%%%%%%%

	% Don't change these lines
	\bsp	% typesetting comment
	\label{lastpage}
\end{document}